# Intertwined electron-nuclear motion in frustrated double ionization in driven heteronuclear molecules

**A. Vilà**

Department of Physics and Astronomy, University College London, Gower Street, London WC1E 6BT, United Kingdom

**Z. Jinzhen**

Physics Department, Ludwig Maximilians Universität, D-80333 Munich, Germany

**A. Scrinzi**

Physics Department, Ludwig Maximilians Universität, D-80333 Munich, Germany

**A. Emmanouilidou**

Department of Physics and Astronomy, University College London, Gower Street, London WC1E 6BT, United Kingdom

**Abstract.** We study frustrated double ionization in a strongly-driven heteronuclear molecule $HeH^+$ and compare with $H_2$. We compute the probability distribution of the sum of the final kinetic energies of the nuclei for strongly-driven $HeH^+$. We find that this distribution has more than one peak for strongly-driven $HeH^+$, a feature we do not find to be present for strongly-driven $H_2$. Moreover, we compute the probability distribution of the n quantum number of frustrated double ionization. We find that this distribution has several peaks for strongly-driven $HeH^+$, while the respective distribution has one main peak and a "shoulder" at lower n quantum numbers for strongly-driven $H_2$. Surprisingly, we find this feature to be a clear signature of the intertwined electron-nuclear motion.

PACS numbers: 33.80.Rv, 34.80.Gs, 42.50.Hz

## 1. Introduction

Frustrated double ionization (FDI) is a major process in the nonlinear response of molecules driven by intense laser fields. In frustrated ionization an electron first tunnel-ionizes in the driving laser field. Then, due to the electric field of the laser pulse, it is recaptured by the parent ion in a Rydberg state [1]. In FDI an electron escapes and another one occupies a Rydberg state at the end of the laser field. Another reason for the experimental and theoretical interest in the FDI process is that FDI is a candidate for the inversion of $N_2$ in free-space air lasing [2]. Other nonlinear phenomena that take place in molecules driven by intense near-infrared (near-IR) laser fields [3] include bond-softening and above threshold dissociation [4, 5], molecular non-sequential double ionization [6–9] and enhanced ionization [9–14]. A number of experimental studies have addressed FDI in the context of $H_2$ [15], $N_2$ [16], Ar dimers [17] and the triatomic molecules $D_3^+$ and $H_3^+$ [18–20]. Frustrated double ionization has also been addressed in theoretical studies in the context of strongly-driven $H_2$ [21] and $D_3^+$ [22] and has been found to account roughly for 10% of all ionization events [15, 21, 22]. Two pathways of FDI have been identified. Electron-electron correlation has been found to play a significant role only for one of the two pathways.

Here, we study FDI for a two-electron heteronuclear diatomic molecule, namely, $HeH^+$. Tracing the dynamics of the electrons and the nuclei at the same time poses a challenge for theory. To overcome this difficulty classical models have been developed which are faster compared to quantum techniques and provide significant insights into the multi-electron dynamics and the interplay of electron-nuclear motion. We have formulated such a 3D semi-classical model in the context of strongly-driven $H_2$ [23] and $D_3^+$ [22] in order to describe FDI through Coulomb explosion. Our 3D semi-classical model accounts both for the motion of the electrons and the nuclei. Our results for both molecules were in good agreement with experimental results [1, 18, 20]. Previous theoretical studies of $HeH^+$ have addressed non-sequential double ionization, by solving the one-dimensional (1D) time-dependent Shrödinger equation (TDSE) [24] as well as by using a 3D soft-core classical ensemble [25], and enhanced ionization by solving the 3D-TDSE [26]. However, in these previous studies the nuclei were kept fixed.

In this work, fully accounting for electron and nuclear motion, we obtain the probability distributions of the sum of the final kinetic energies of the nuclei of FDI for strongly-driven $HeH^+$ and compare with $H_2$. We also compute the probability distribution of the n quantum number of FDI for strongly-driven $HeH^+$ and compare with the respective distribution for $H_2$. We find a very interesting feature of the distribution of the n quantum number, namely, the presence of several peaks. We show that these peaks are signatures of the intertwined electron-nuclear dynamics.

## 2. Method

We employ a linearly polarized laser field that is of the following form:

$$\mathbf{E}(t) = E_0 f(t) cos(\omega t)\hat{z}$$
$$f(t) = exp\left(-2ln2\left(\frac{t}{\tau_{FWHM}}\right)^2\right), \qquad (1)$$

with $\omega = 0.057$ a.u. for commonly used Ti:sapphire lasers at 800 nm and $\tau_{FWHM} = 40$ fs the full-width-half-maximum. The strength of the laser field $E_0$ is equal to 0.2 a.u. for strongly-driven HeH$^+$ and equal to 0.0564 a.u. for strongly-driven H$_2$. Both electric field strengths are within the below-the-barrier ionization regime. The electric field strength for our studies of strongly-driven HeH$^+$ is chosen such that it is close to the threshold field strength of 0.272 a.u. for over-the-barrier ionization. The electric field strength for H$_2$ is chosen such that it has the same percentage difference from the threshold field strength of 0.0768 a.u. for over-the-barrier ionization as HeH$^+$ has from 0.272 a.u. The strengths are chosen such that the probabilities for frustrated double ionization are roughly of the same order of magnitude for both diatomic molecules.

To formulate the initial state of the two electrons, we assume that one electron (electron 1) tunnel-ionizes at time $t_0$ in the field-lowered Coulomb potential. For this quantum-mechanical step, we compute the ionization rate for HeH$^+$ using a quantum-mechanical calculation. Specifically, alignment dependent tunnel-ionization rates for the initial state of HeH$^+$ at a fixed internuclear distance of 1.46 a.u. were obtained using the haCC (hybrid anti-symmetrized Coupled Channels) method described in [27]. In haCC, the system is represented in a hybrid basis where the ground state of HeH$^+$ and the energetically lowest few HeH$^{2+}$ states are drawn from the standard quantum chemistry package COLUMBUS [28], while a purely numerical basis is used to represent the tunneling electron. Anti-symmetrization is fully enforced. Tunnel ionization rates were computed using exterior complex scaling. For details of the method and discussion of its accuracy, see Ref. [27]. For H$_2$ we compute the ionization rate using a semi-classical formula [29]. In our computations, the time that electron 1 tunnel-ionizes, $t_0$, is selected according to the ionization rate in the time interval the laser field is switched on. To efficiently obtain $t_0$ we use importance sampling [30] with the ionization rate as the importance sampling distribution. For electron 1, the velocity component that is transverse to the laser field is given by a Gaussian [31] and the component that is parallel is set equal to zero. We note that for each field strength, which corresponds to a certain $t_0$, the ionization rate is higher when electron 1 tunnel-ionizes from the H$^+$ side rather than the He$^{2+}$ side. For the maximum field strength we consider, i.e. 0.2 a.u., the ionization rate is roughly three times higher when electron 1 tunnel-ionizes from the H$^+$ rather than from the He$^{2+}$ side.

The initial state of the initially bound electron (electron 2) is described by a microcanonical distribution [32] of HeH$^{2+}$ at the internuclear distance of HeH$^+$. In Fig. 1, we compare the microcanonical and quantum mechanical probability density of the electron 2 position on the x-z plane for all values of the y-component for $HeH^{2+}$. We find that the microcanonical distribution overestimates the probability for electron 2 to be around the $H^+$ nucleus.

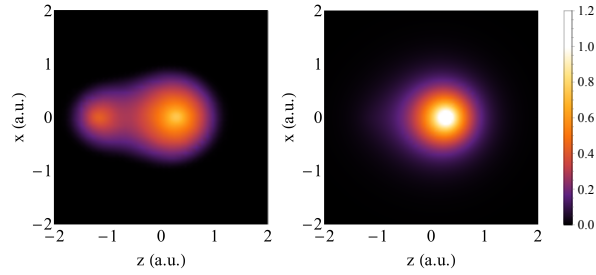

**Figure 1.** Left panel: the microcanonical probability density of the electron 2 position on the x-z plane for all values of the y-component; Right panel: the quantum mechanical probability density of the electron 2 position on the x-z plane, integrating over all values of the y-component.

Another quantum mechanical aspect of our 3D model is tunneling of each electron during the propagation with a probability given by the Wentzel-Kramers-Brillouin approximation [21, 23]. This aspect is essential to accurately describe the enhanced ionization process (EI) [7, 33]. In EI, at a critical distance of the nuclei, a double potential well is formed such that it is easier for an electron bound to the higher potential well to tunnel to the lower potential well and subsequently ionize. We refer to the last time an electron tunnels during the propagation as $t_{tun}$. Our 3D semiclassical model also accounts for events where an electron does not tunnel, however, for intermediate intensities in most FDI and doubly-ionizing events the electron/electrons escape after tunneling [22]. The time propagation is classical, starting from time $t_0$. We solve the classical equations of motion for the Hamiltonian of the strongly-driven four-body system, while fully accounting for the Coulomb singularities [23].

Note that in what follows we refer to an event where an electron tunnels through a potential barrier as a tunnel-ionization event. The reason is that eventhough the electron tunnels through a potential barrier it can be that the final, in the asymptotic time limit, energy of this electron is not positive and thus the electron does not ionize. In our computations the asymptotic time limit is several millions in atomic units. An electron that has positive final energy is referred to as an escaping electron. After propagating the trajectories to the asymptotic time limit, we record all ionization events, double and single ionization events as well as events where both electrons remain bound. Our 3D semiclassical model accurately describes the processes that take place through Coulomb explosion.

A subset of the single ionization events are FDI events where we select trajectories that produce either $H^+$, a free electron and $He^{+*}$ (where $*$ denotes that the electron is in a $n > 1$ quantum state) or $H^*$, a free electron and $He^{2+}$. To identify the trajectories when the electron is captured in an excited state, we first find the classical principal number $n_c = Z_i/\sqrt{2|\epsilon|}$, where $\epsilon$ is the total energy of the trapped electron and $Z_i$ is either 1 or 2 depending on whether the electron remains attached to $H^*$ or $He^{+*}$, respectively; i= 1, 2 for each of the two nuclei. We, next, assign a quantum number so that the following criterion, which is derived in [34], is satisfied:

$$[(n-1)(n-1/2)n]^{1/3} \leq n_c \leq [n(n+1/2)(n+1)]^{1/3}. \qquad (2)$$

## 3. Percentage of FDI and of pathways A and B of FDI

We find that two pathways lead to frustrated double ionization, A and B, as previously found for strongly-driven $H_2$ [21]. The difference between the two FDI pathways lies in how fast the ionizing electron escapes following the turn on of the laser field [21]. In pathway A, electron 1 tunnel-ionizes and escapes early on. Electron 2 gains energy from an EI-like process and tunnel-ionizes. It does not have enough drift energy to escape when the laser field is turned off and finally it occupies a Rydberg state, $H^*$ or $He^{+*}$. In pathway B, electron 1 tunnel-ionizes and quivers in the laser field returning to the core. Electron 2 gains energy from both an EI-like process and the returning electron 1 and tunnel-ionizes after a few periods of the laser field. When the laser field is turned off, electron 1 does not have enough energy to escape and remains bound in a Rydberg state. It follows that electron-electron correlation is more pronounced in pathway B [21, 35].

For $HeH^+$ we find that the probability of FDI (n ≥ 2) out of all ionization events is 2.43%, with 1.34% being the probability for electron attachment to the n = 2 state. This is unlike strongly-driven $H_2$, where the probability of FDI is much higher and equal to 7.45%, while the probability for electron attachment to the n = 2 state is only 0.17%. In Table 1, we show that the n = 2 state for strongly-driven $HeH^+$ is mainly populated via pathway A when electron 2 is attached to the $He^{+*}$ ion. In particular, our analysis shows that, for most of these latter events (85%), electron 2, just before it gets attached to the $He^{2+}$ nucleus, tunnels at time $t_{tun}$ from the potential well corresponding to $He^+$ to the potential well corresponding to $He^{2+}$. Moreover, we find that, before tunneling, electron 2 has energy corresponding to the n = 1 state of the H atom, however, this energy corresponds to the n = 2 state of the $He^+$ atom. In pathway B, the probability for electron 1 attachment to n = 2 states is much smaller. In pathway B, electron 1, after being accelerated in the field, when it returns to the molecular ion has a larger energy than electron 2 and a smaller chance of getting finally attached to a low n quantum number state. We note that the probability of double ionization is 14.8% and 33.25% for $HeH^+$ and $H_2$, respectively.

We find (not shown) that the distribution of the sum of the final kinetic energies of the nuclei for strongly-driven $HeH^+$ for FDI events where an electron finally remains attached to the n = 2 state peaks at small energies. Therefore, most probably, FDI events with n = 2 can not be distinguished from single ionization events resulting from bond-softening. The latter events are not accounted for by our computations. For this reason, in what follows we focus on FDI events with n > 2 for strongly-driven $HeH^+$.

We also find that the probability for FDI for electron attachment to n > 2 states is roughly three times higher through electron-electron correlation in pathway B than through pathway A. Moreover, for pathway A (B) of FDI we find that the probability for electron 2 (electron 1) to get attached to the n > 2 states of $He^{+*}$ is three times higher than the probability for electron attachment to the n > 2 states of $H^{+*}$. The higher probability for electron attachment to $He^{+*}$ compared to $H^*$ is consistent with the higher charge of $He^{2+}$ compared to the charge of $H^+$.

Table 1. The probability of FDI for pathways A and B for strongly-driven molecules $HeH^+$ and $H_2$.

|  | PA | H | H $n=2$ | H $n>2$ | He | He $n=2$ | He $n>2$ |
|---|---|---|---|---|---|---|---|
| $HeH^+$ (%) | 1.61 | 0.072 | 0.003 | 0.069 | 1.54 | 1.33 | 0.21 |
| $H_2$ (%) | 2.77 | 2.77 | 0.12 | 2.65 | - | - | - |
|  | PB | H | H $n=2$ | H $n>2$ | He | He $n=2$ | He $n>2$ |
| $HeH^+$ (%) | 0.82 | 0.2 | 0 | 0.2 | 0.62 | 0.01 | 0.61 |
| $H_2$ (%) | 4.68 | 4.68 | 0.05 | 4.64 | - | - | - |

## 4. Probability distributions of the sum of the final kinetic energies of the nuclei

In Fig. 2, we plot the probability distribution of the sum of the final kinetic energies of the two nuclei (KER) for the FDI process. That is, we plot the KER of FDI when one of the two electrons finally remains attached either to the $H^*$ or the $He^{+*}$ final fragment with n > 2 for strongly-driven $HeH^+$. We find that there is one main peak and a secondary one in the distribution of the KER of FDI. To identify the origin of these peaks for $HeH^+$, we consider the contribution to the distribution of KER of each pathway of FDI separately and compare with $H_2$. Comparing Fig. 3(a) with (c) for pathway A and Fig. 3(b) with (d) for pathway B, we find that the distributions of KER for both pathways of FDI peak at much higher energies for $HeH^+$ than for $H_2$, i.e., at 16 eV in the former case compared to roughly 6 eV in the latter case. This main peak is denoted as E1 for pathway A and as E1′ for pathway B. To understand the higher energy values of the E1 and the E1′ peaks for HeH$^+$ compared to H$_2$, we compute the distribution of the inter-nuclear distance at the time $t_{\text{tun}}$ electron 2 tunnel-ionizes for both molecules (not shown). We find that the most probable inter-nuclear distance at the time electron 2 tunnel-ionizes is around $R_{\text{tun}}$ =4.3 a.u. for both pathways and for both molecules. This distance is in accord with the internuclear distance at the the time the EI process takes place [10]. Assuming that Coulomb explosion of the nuclei takes place at the time electron 2 tunnel-ionizes, the most probable value of the sum of the final kinetic energies of the nuclei should be roughly given by $Z_1Z_2/R_{\text{tun}}$ which is equal to 6.3 for H$_2$ and 12.7 eV for HeH$^+$. However, the actual values of the location of the peaks of the distributions of KER differ from the predicted ones, more so for HeH$^+$. This difference can be accounted for if at the time of tunnel-ionization of electron 2 the distribution of the sum of the kinetic energies of the nuclei peaks at a higher energy for HeH$^+$ than for H$_2$. This is indeed the case. Comparing Fig. 4(a) for n > 2 with (c) for pathway A and Fig. 4(b) for n > 2 with (d) for pathway B, we find that the most probable value of the sum of the kinetic energies at the time electron 2 tunnel-ionizes is 3.5 eV for HeH$^+$ versus 0.8 eV for H$_2$ for pathway A and 4.6 eV for HeH$^+$ versus 1.4 eV for H$_2$ for pathway B.

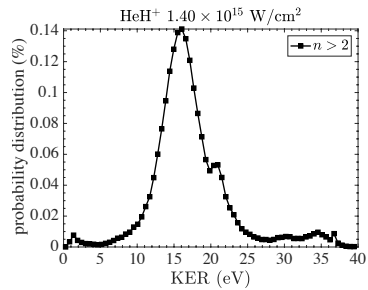

**Figure 2.** Probability distributions of the KER of FDI for n > 2, for strongly-driven $HeH^+$.

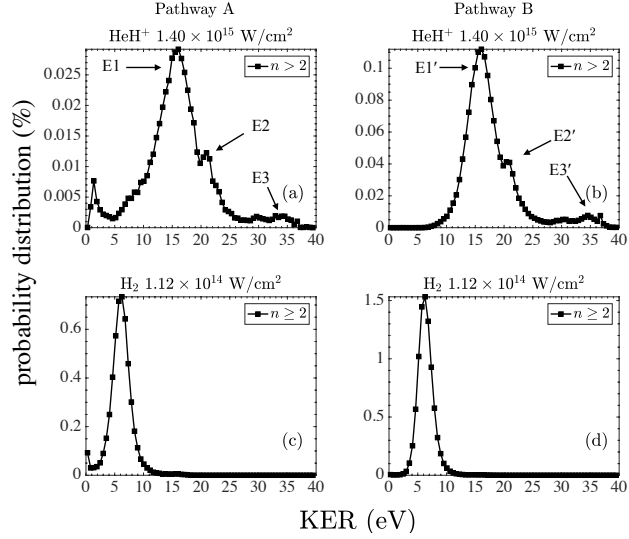

**Figure 3.** Probability distributions of the KER for pathway A (left panels) and pathway B (right panels) of FDI for strongly-driven HeH$^+$, (a) and (b), and for H$_2$, (c) and (d).

Still focusing on electron attachment to n > 2 states for strongly-driven HeH$^+$, comparing Fig. 3(a) with (c) for pathway A and Fig. 3(b) with (d) for pathway B, we find that, unlike H$_2$, for HeH$^+$ the distributions of the KER for both pathways have a second smaller and a third much smaller peak roughly at 21 eV and 35 eV, respectively.

The second and third peaks are denoted as E2 and E3 for pathway A, and as E2′ and E3′ for pathway B. To understand the origin of these peaks in HeH$^+$, we plot in Fig. 5 and Fig. 6 for pathway A and B, respectively, the double differential probability of the KER as a function of the time electron 2 tunnel-ionizes with respect to the initial time, i.e $t_{\text{tun}}$- $t_0$, and as a function of the initial coordinate of electron 2 along the inter-nuclear axis. We find that the three different peaks correspond to different initial conditions and different times of tunnel-ionization of electron 2. For strongly-driven H$_2$, Fig. 5(d) and Fig. 6(d) for pathway A and B, respectively, show that the single peak in the distributions of KER corresponds to events with electron 2 being initially located mainly between the two nuclei. This is also the case for the main peaks E1 and E1′ of the distributions of KER for strongly-driven HeH$^+$, see Fig. 5(c) and Fig. 6(c) for pathway A and B, respectively. However, Fig. 5(c) and Fig. 6(c) also show that for strongly-driven HeH$^+$ there is a considerable probability for FDI events with electron 2 being initially localized around the nuclei. We find that these later FDI events give rise to the two higher energy peaks of the distributions of KER for HeH$^+$. Specifically, the E2 and E2′ peaks originate from electron 2 being localized mostly around the He$^{2+}$ nucleus. Moreover, for these FDI events electron 2 tunnel-ionizes earlier in time compared to the FDI events that give rise to the main peak of the distributions of KER, see Fig. 5(a) and Fig. 6(a) for pathway A and B, respectively. Comparing Fig. 5(a) with (b) for pathway A and Fig. 6(a) with (b) for pathway B, we find that on average electron 2 tunnel-ionizes at larger times for strongly-driven H$_2$ compared to HeH$^+$. Thus, the additional peaks in the distributions of the KER for both pathways of FDI for HeH$^+$ are due to events with electron 2 being initially localized around the nuclei and tunnel-ionizing early on during the time propagation. It could be the case that experimentally only one peak will be observed with an energy greater than the E1 peak and smaller than the E2 peak. The reason is that the microcanonical distribution we employ to describe the initial state of electron 2 overestimates events where electron 2 is localized around H$^+$ and in between the nuclei, see Fig. 1.

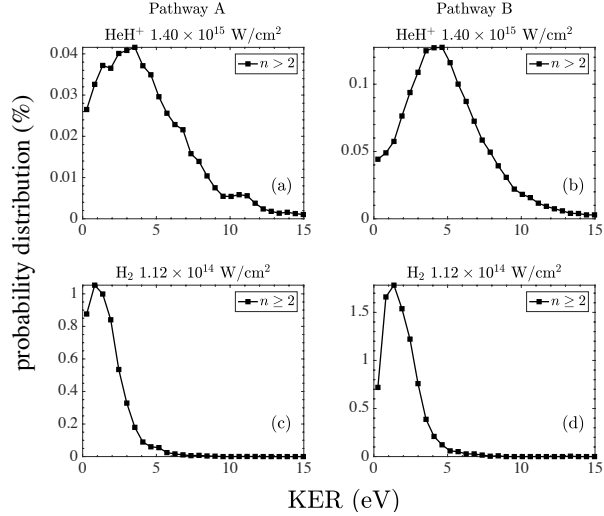

**Figure 4.** Probability distributions of the sum of the kinetic energies of the nuclei at the time electron 2 tunnel-ionizes for pathway A (left panels) and B (right panels) of FDI for strongly-driven HeH$^+$, (a) and (b), and for H$_2$, (c) and (d).

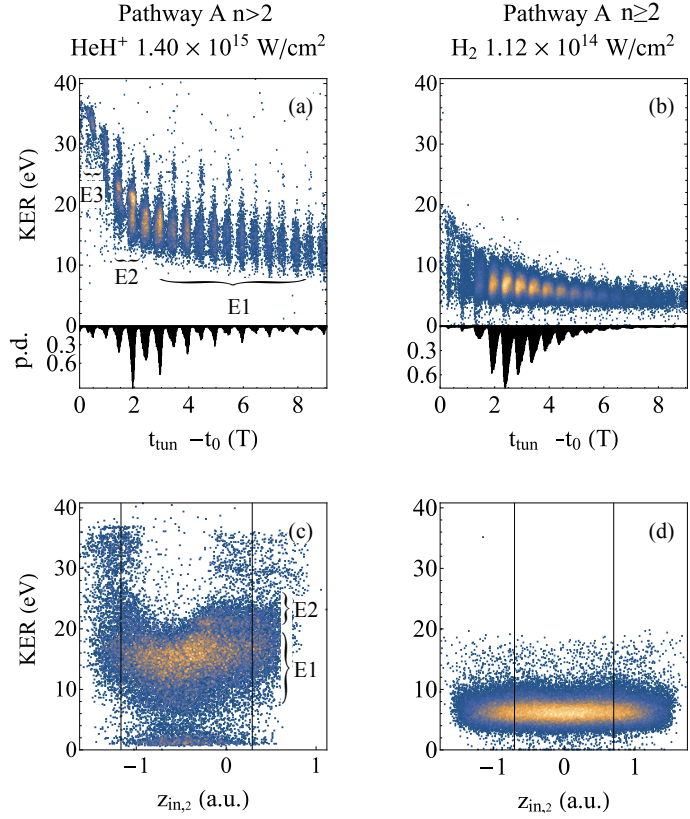

**Figure 5.** Top panels: double differential probability of the KER for pathway A of FDI as a function of the time electron 2 tunnel-ionizes with respect to $t_0$ (a) for strongly-driven HeH$^+$ with n > 2, and (b) for strongly-driven H$_2$; Bottom panels: double differential probability of the KER for pathway A of FDI as a function of the initial coordinate of electron 2 along the inter-nuclear axis (c) for strongly-driven HeH$^+$ with n > 2, and (d) for strongly-driven H$_2$. For the top panels we also plot the probability distributions, i.e., p.d, for the time electron 2 tunnel-ionizes with respect to $t_0$.

## 5. Signatures of interplay of electron-nuclear motion on the distribution of the n quantum number

In Fig. 7, we plot the probability distributions of the n quantum number for pathways A and B of FDI for electron attachment to H$^*$ and He$^{+*}$ for strongly-driven HeH$^+$. We find that these distributions have more than one main peak. These peaks are particularly pronounced for pathway B of FDI and for electron attachment to He$^{+*}$, see Fig. 7(b). In Fig. 8, for strongly-driven H$_2$, we find that the distributions for the n quantum number have only one main peak and a broad "shoulder" at smaller values of the n quantum number. This "shoulder" is particularly pronounced for pathway A of FDI.

In what follows, we identify the origin of these peaks and "shoulder" and explain why for strongly-driven HeH$^+$ and for electron attachment to He$^{+*}$ the distribution of the n quantum number has several peaks. Very interestingly, we find that, regarding the distribution of the n quantum number, the presence of several peaks for electron attachment to He$^{+*}$ for strongly-driven HeH$^+$ as well as the broad "shoulder" for strongly-driven H$_2$ are signatures of the intertwined electron-nuclear motion. Indeed, in our analysis of FDI events, following tunnel-ionization of electron 2, that is, after time $t_{tun}$ we identify the number of times, Nmax, the electron that finally remains bound in a Rydberg state goes back and forth between the nucleus the electron finally gets attached to and the other nucleus. Namely, for electron attachment to He$^{+*}$ for pathway A and B we identify the number of maxima in time of $\frac{|\tilde{r}_{He^{2+}} - \tilde{r}_i|}{|\tilde{r}_{H^+} - \tilde{r}_i|}$ when $E_{pot,i}^{H^+} < E_{pot,i}^{He^{2+}}$, with $E_{pot,i}^{H^+}$ ($E_{pot,i}^{He^{2+}}$) being the potential energy of electron i with respect to H$^+$ (He$^{2+}$); $\vec{r}_{H^+}, \vec{r}_{He^{2+}}, \vec{r}_i$ are the position vectors of the nuclei H$^+$ and He$^{2+}$ and of electron i, respectively, while i is equal to 2 for pathway A and 1 for pathway B. Similarly, for electron attachment to the H$^*$ atom for pathway A and B we identify the number of maxima in time of $\frac{|\tilde{r}_{H^+} - \tilde{r}_i|}{|\tilde{r}_{He^{2+}} - \tilde{r}_i|}$ when $E_{pot,i}^{He^{2+}} < E_{pot,i}^{H^+}$. The above definition determines approximately Nmax since it depends on the position of the electron that finally remains in a Rydberg state at the time $t_{tun}$ electron 2 tunnel-ionizes. For instance, for pathway A and for electron attachment to $\text{He}^{+*}$, if electron 2, following tunnel-ionization, is located closer to $\text{H}^+$ with $\text{E}^{\text{H}^+}_{\text{pot},2} < \text{E}^{\text{He}^{2+}}_{\text{pot},2}$ and if then electron 2 proceeds to approach $\text{He}^{2+}$, we still register this as a maximum. So there is an ambiguity of one count in our definition of Nmax, which, however, does not affect the conclusions we draw in what follows. We find that there is no such ambiguity for strongly-driven $\text{HeH}^+$ for pathway B and electron 1 getting attached to $\text{He}^{+*}$, since electron 1, following tunnel-ionization of electron 2, is mostly located on the $\text{He}^{2+}$ side. We also find that for strongly-driven $\text{H}_2$, electron 2 (1) for pathway A (B) has the same probability to be positioned closer to either nucleus at the time that electron 2 tunnel ionizes.

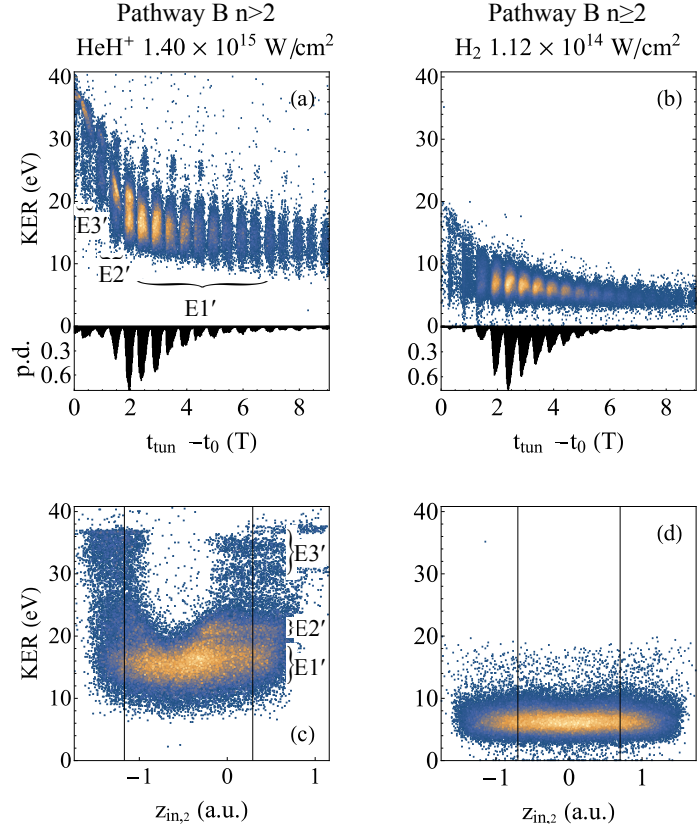

**Figure 6.** Top panels: double differential probability of the KER for pathway B of FDI as a function of the time electron 2 tunnel-ionizes with respect to $t_0$ (a) for strongly driven HeH$^+$, and (b) for strongly-driven H$_2$; Bottom panels: double differential probability of the KER for pathway B of FDI as a function of the initial coordinate of electron 2 along the inter-nuclear axis (c) for strongly driven HeH$^+$, and (d) for strongly-driven H$_2$. For the top panels we also plot the probability distributions, i.e., p.d, for the time electron 2 tunnel-ionizes with respect to $t_0$.

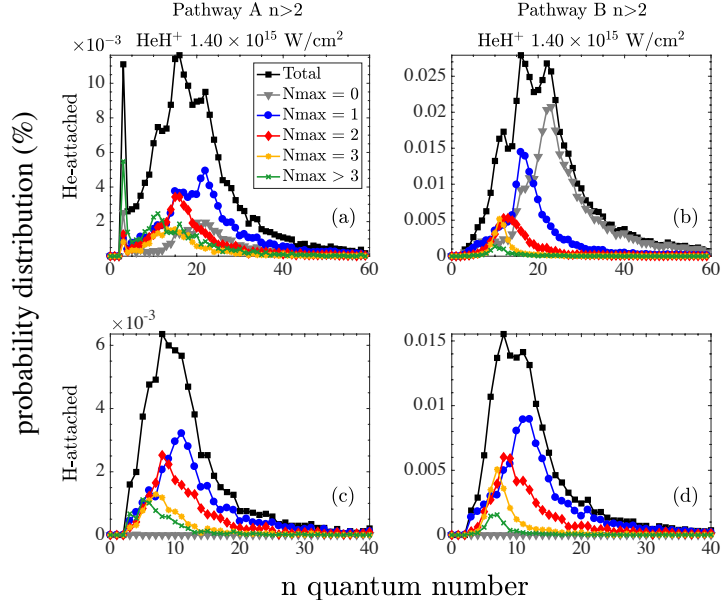

**Figure 7.** Top panels: the probability distributions of the n quantum number (a) for pathway A and (b) for pathway B of FDI for electron attachment to $\text{He}^{+*}$ for strongly-driven $\text{HeH}^+$ for all Nmax numbers (black line), for Nmax = 0 (grey line), for Nmax = 1 (blue line), for Nmax = 2 (red line), for Nmax = 3 (orange line), and for Nmax $\geq$ 3 (green line). Bottom panels similar to top panels for electron attachment to $\text{H}^*$.

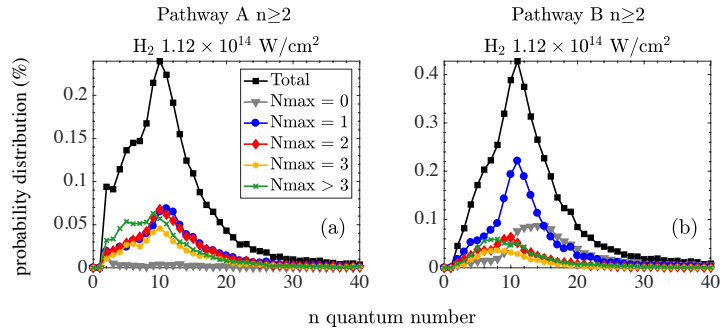

**Figure 8.** Similar to bottom panels in Fig. 7 for strongly-driven $\text{H}_2$.

Table 2 shows that for strongly-driven $\text{H}_2$ the electron that finally remains attached to a Rydberg state of an $\text{H}^*$ atom approaches the other $\text{H}^+$ nucleus a larger number of times for pathway A compared to pathway B. This is consistent with electron 1, which finally remains bound in pathway B, having more energy to start with than electron 2, which finally remains bound in pathway A. Indeed, initially, electron 2 is bound while electron 1 tunnel-ionizes. Next, in Fig. 7 and Fig. 8, we also plot for each different value of Nmax the corresponding probability distribution, Nmax-distribution, of the n quantum number for each pathway of FDI for strongly-driven $HeH^+$ and $H_2$. We find that a Nmax-distribution of the n quantum number peaks at a higher n number the smaller Nmax is. This is particularly pronounced for electron 1 attachment to $He^{+*}$ for pathway B of FDI for strongly-driven $HeH^+$, see Fig. 7(b). Indeed, in the latter case the distributions of the n quantum number with Nmax = 0 and Nmax = 1 peak at n = 22 and n = 16, respectively, giving rising to the main two peaks of the distribution of the n quantum number when all Nmax values are included. Moreover, the Nmax-distributions with Nmax $\geq$ 2 peak at n = 10 − 12 and give rise to the third peak of the distribution of the n quantum number when all Nmax values are included. Similar results hold for electron 2 attachment to $He^{+*}$ for pathway A of FDI for strongly-driven $HeH^+$, see Fig. 7(a). In addition, we see in Fig. 7(c) and (d) that for electron 2 (1) attachment to $H^*$ for pathway A (B) for strongly-driven $HeH^+$, the two main peaks at n = 8, 11 of the distribution of the n quantum number when all Nmax values are included are accounted for by the Nmax-distributions of the n quantum number with Nmax = 2 and Nmax = 1, respectively. This correspondence between Nmax and a different peak in the distribution of the n quantum number is not as clear for strongly-driven $H_2$ compared to $HeH^+$, see Fig. 8. However, even for strongly-driven $H_2$ for pathway B the big "shoulder" for small n quantum numbers corresponds to Nmax $\geq$ 2. We note that we have checked and the probability distribution of the initial position of electron 2 along the inter-nuclear axis is the same for each Nmax labeled FDI events. This means that the microcanonical distribution for electron 2 is not responsible for the increased number of peaks in the distribution of the n quantum number for strongly-driven $HeH^+$.

Table 2. Percentage of FDI events where the electron that finally remains attached to one of the two nuclei for pathway A (PA) and B (PB) approaches the other nucleus $0,1,2,3, > 3$ times for strongly-driven $HeH^+$ and $H_2$.

|  | % 0 peaks | % 1 peaks | % 2 peaks | % 3 peaks | % >3 peaks |
|---|---|---|---|---|---|
| He-PA | 16.1 | 37.0 | 19.5 | 9.9 | 17.4 |
| H-PA | 3.9 | 44.8 | 29.0 | 12.5 | 9.7 |
| He-PB | 59.1 | 26.0 | 9.3 | 4.2 | 1.5 |
| H-PB | 0.6 | 53.7 | 29.1 | 12.8 | 3.8 |
| $H_2$ PA | 2.6 | 27.3 | 26.3 | 17.3 | 26.3 |
| $H_2$ PB | 24.8 | 40.2 | 12.9 | 8.2 | 13.9 |

We find that the electron dynamics is intertwined with the dynamics of the nuclei mostly for pathway B. Indeed, in Fig. 9 we plot for pathway B and for attachment to $He^{+*}$ the probability distribution of the inter-nuclear distance of the two nuclei at the time of the occurrence of the single peak in the Nmax-distribution with Nmax=1 and at the time of the occurrence of each of the two peaks in the Nmax-distribution with

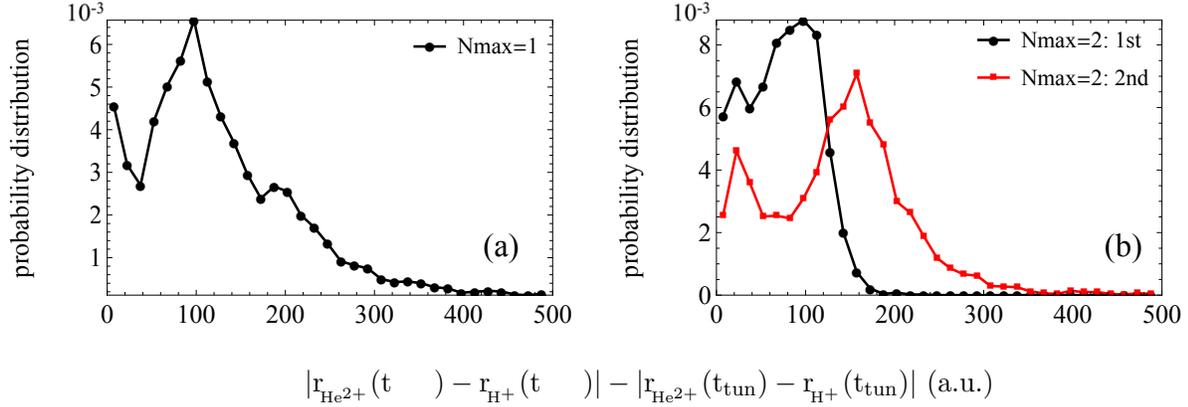

$|r_{He^{2+}}(t) - r_{H^+}(t)| - |r_{He^{2+}}(t_{tun}) - r_{H^+}(t_{tun})|$ (a.u.)

**Figure 9.** Probability distribution of the inter-nuclear distance at the time when the peak Nmax is registered with respect to the inter-nuclear distance at the time $t_{tun}$ electron 2 tunnel-ionizes for Nmax = 1 (a) and Nmax = 2 (b) for pathway B for attachment to $He^{+*}$.

Nmax=2. These distributions are plotted with respect to the inter-nuclear distance at the time $t_{tun}$ electron 2 tunnel-ionizes, with the latter distribution having a peak around 4.25 a.u. We find that, following time $t_{tun}$, at the time electron 1 first approaches the $H^+$ nucleus the nuclei have already moved significantly apart with their most probable inter-nuclear distance being around 100 a.u., see Fig. 9(a) and Fig. 9(b) black line. We also find (not shown) that, following time $t_{tun}$, the most probable time electron 1 first approaches the $H^+$ nucleus is 31 laser field cycles. By the time electron 1 approaches for a second time the $H^+$ nucleus the nuclei have moved even further apart with the most probable inter-nuclear distance being around 170 a.u., see Fig. 9(b) red line. We also find (not shown) that, following time $t_{tun}$, the most probable time electron 1 approaches for a second time the $H^+$ nucleus is 52 laser field cycles.

Next, we explain why the correspondence between Nmax and a peak of the distribution of n quantum numbers is more pronounced for electron attachment to $He^{+*}$ versus $H^*$. For hydrogenic atoms, the difference between two energy levels corresponding to subsequent n quantum numbers is larger for the higher charged nucleus. In our case this energy difference is four times larger for the He atom. Moreover, we find that the energy of the electron that finally remains bound in a Rydberg state is similar both for attachment to $He^{+*}$ and to $H^*$. Given the above, each small interval of this final energy of the electron that remains finally bound encompasses more n quantum numbers for attachment to $H^*$ than for attachment to $He^*$. Therefore, the intertwined dynamics of the finally bound electron with the nuclei, which is directly related to the final energy this electron has, is imprinted with a higher "resolution", i.e., smaller spread over n quantum numbers for attachment to the higher charged ion. This is consistent with the higher number of peaks in the distribution of the n quantum number for pathway A and B of FDI for final attachment to $He^{+*}$ for $HeH^+$, see Fig. 7(a) and Fig. 7(b), compared to the distribution of the n quantum number for pathway A and B of FDI for final attachment to $H^*$, see Fig. 7(a) and Fig. 7(b) for $HeH^+$ and Fig. 8(a) and Fig. 8(b) for $H_2$.

## 6. Conclusions

We have studied frustrated double ionization for strongly-driven $HeH^+$. We have computed the probability distributions of the KER of FDI for strongly-driven $HeH^+$ and $H_2$. We have found that, while there is one peak in the distribution of KER for strongly-driven $H_2$, there is a main peak and a couple of secondary ones in the distribution of KER for strongly-driven $HeH^+$. We have shown that these peaks for strongly-driven $HeH^+$ probe different initial conditions of the position of the initially bound electron. Namely, the secondary peaks correspond to the initially bound electron being localized around the nuclei and mostly around the $He^{2+}$ nucleus. Very interestingly, we have shown that the probability distribution of the n quantum number has several peaks for strongly-driven $HeH^+$, while the respective distribution has only one main peak and a "shoulder" for lower n quantum number for strongly-driven $H_2$. We have found that this feature is related to the interplay of the electron dynamics with the dynamics of the nuclei. To show that this is the case, we have computed the number of times, Nmax, the electron that finally remains attached to a Rydberg state goes back and forth between the two nuclei following tunnel-ionization of the initially bound electron. We have shown that there is a correspondence between Nmax and the location of the peak of the probability distribution of the n-quantum number that is computed using the Nmax labelled FDI events. Namely, a high Nmax corresponds to a distribution of the n quantum number that peaks at small values of n. Thus, we have shown that the presence of several peaks in the distribution of the n quantum number is a clear signature of the intertwined electron-nuclear motion. We conjecture that this feature will be more pronounced for nuclei with a higher charge than $He^{2+}$. However, the higher the charge of the nucleus is the less likely is the FDI process and, thus, the more difficult it is to observe such features.

*Acknowledgments.* A.E. acknowledges the EPSRC grant no. J0171831 and the use of the computational resources of Legion at UCL. J. Zh. was supported by the DFG Priority Programme 1840, QUTIF.